\documentclass{jpsj-suppl}
\usepackage{times}
\usepackage{color}

\title{$^{7}$Li NMR Studies of LiCrO$_{2}$}

\author{Yutaka ITOH\thanks{E-mail address: yitoh@cc.kyoto-su.ac.jp} 
}
\inst{Department of Physics, Graduate School of Science, Kyoto Sangyo 
University, Kamigamo-Motoyama, Kika-ku, Kyoto 603-8555, Japan\\
}

\recdate{September 30, 2013}

\abst{
We report on $^{7}$Li NMR studies of a spin $S$ = 3/2 triangular lattice antiferromagnet LiCrO$_2$ (N\'{e}el temperature $T_\mathrm{N}$ = 62 K) in the paramagnetic state by using the free-induction decay of $^{7}$Li nuclear magnetization. We observed critical divergence of the $^{7}$Li nuclear spin-lattice relaxation rate 1/$T_1$ near $T_\mathrm{N}$, a narrow critical region, and a critical exponent $w$ = 0.45 from a fit of 1/$T_1$ $\propto$ ($T/T_\mathrm{N}$ - 1)$^{-w}$. Although spin frustration effects have been explored for this system, the dynamical critical phenomena suggest that LiCrO$_2$ in the critical region is a poor low dimensional antiferromagnetic system.  
}

\kword{spin frustration effect, NMR, LiCrO$_2$}

\begin{document}
\maketitle

\section{Introduction}

LiCrO$_2$ is a quais-two dimensional triangular lattice Heisenberg antiferromagnet with a N\'{e}el temperature $T_\mathrm{N}$ $\approx$ 62 K. The crystal structure is an ordered rock salt structure, where the Cr$^{3+}$ ion carries a local moment of $S$ = 3/2 on the triangular lattice. Non Curie-Weiss behavior of the uniform spin susceptibility below about 450 K suggests a low dimensional exchange network and a possible spin frustration effect. The Weiss temperature $\Theta$ is estimated to be -700 K from the Curie-Weiss susceptibility fit above about 450 K~\cite{XT}. The spin frustration effects on the paramagnetic state and the magnetic ordered state have been explored by using ESR~\cite{Ajiro}, neutron diffraction~\cite{Kadowaki}, NMR and thermodynamic properties~\cite{NMR} and muon spin rotation ($\mu$SR)~\cite{Sugiyama}. The temperature dependences of the ESR line width~\cite{Ajiro} and $^{7}$Li NMR spin-echo relaxation rate~\cite{NMR} were associated with an exponential increase of thermally excited Z$_2$ vortices (topological defects)~\cite{KM}. However, three dimensional magnetic structure with double-$Q$ 120$^\circ$ ordering vectors has been observed in the neutron diffraction~\cite{Kadowaki} and $\mu$SR~\cite{Sugiyama}. Two dimensional renormalized classical spin correlation also yields an exponential divergence toward $T$ = 0 K in the NMR relaxation rate for a triangular lattice Heisenberg antiferromagnet such as Li$_7$RuO$_6$~\cite{Itoh}. There remains to be solved whether the two dimensional spin correlation predominates in the paramagnetic state of LiCrO$_2$. We have performed a detailed $^{7}$Li NMR experiment using the free induction decay (FID) of $^{7}$Li nuclear magnetization for LiCrO$_2$.

In this paper, we report on $^{7}$Li NMR studies of LiCrO$_2$ polycrystalline samples in the paramagnetic state. We observed critical divergence of the $^{7}$Li nuclear spin-lattice relaxation rate 1/$T_1$ near $T_\mathrm{N}$, a narrow critical region, a critical exponent $w$ = 0.45 by a fit of 1/$T_1$ $\propto$ ($T/T_\mathrm{N}$ - 1)$^{-w}$, and no regimes of the two dimensional renormalized classical spin correlation.  
 
\section{Experiments}

Powder samples of LiCrO$_2$ have been synthesized by a conventional solid-state reaction method. Appropriate amounts of Li$_2$CO$_3$ and Cr$_2$O$_3$ were mixed, palletized and fired at 1150$^\circ$C for 24 hours in air. The products were confirmed to be in a single phase from measurements of the powder X-ray diffraction patterns. A phase-coherent-type pulsed spectrometer was utilized to perform the $^{7}$Li 
NMR (nuclear spin $I$ = 3/2) experiments in an external magnetic field of 1.00 T.    
The NMR frequency spectra were obtained from Fourier transformation of the $^{7}$Li FID signals.   
$^{7}$Li nuclear spin-lattice relaxation curves $^{7}p(t) = 1-F(t)/F(\infty)$ (recovery
curves) were obtained by using an inversion recovery technique as a function of time $t$ after an inversion pulse,  where FID $F(t)$, $F(\infty)[\equiv F(10T_1)]$ and $t$ were recorded.   

\section{Results}

Figure 1(a) shows the Fourier-transformed (FT) spectra of $^7$Li FID signals in LiCrO$_2$ at 77 and 289 K at a reference frequency of 16.5520 MHz. The NMR line of LiCl$aq$ represents the reference frequency at zero shift at 1.00 T. No appreciable change is found in the linewidth of the NMR spectra from 289 K to 77 K. The effects of relocation of the Li ions and the motion, which were suggested by the $\mu$SR studies~\cite{Sugiyama}, could not be found in the NMR spectra of our samples. 

Figure 1(b) shows the recovery curves $^{7}p(t)$ of $^7$Li NMR FID signals as a temperature is decreased. The solid lines are the results from the least-squares fit by a single exponential function  
\begin{equation}
^{7}p(t)=p(0)\rm{exp}{(-\frac{\it t}{^{7}\it T_{\rm 1}})}
\label{eq.2}
\end{equation}
where $p(0)$ and the $^7$Li nuclear spin-lattice relaxation time $^{7}T_1$ are the fit parameters.  

 \begin{figure}[h]
 \begin{center}
 \includegraphics[width=14 cm]{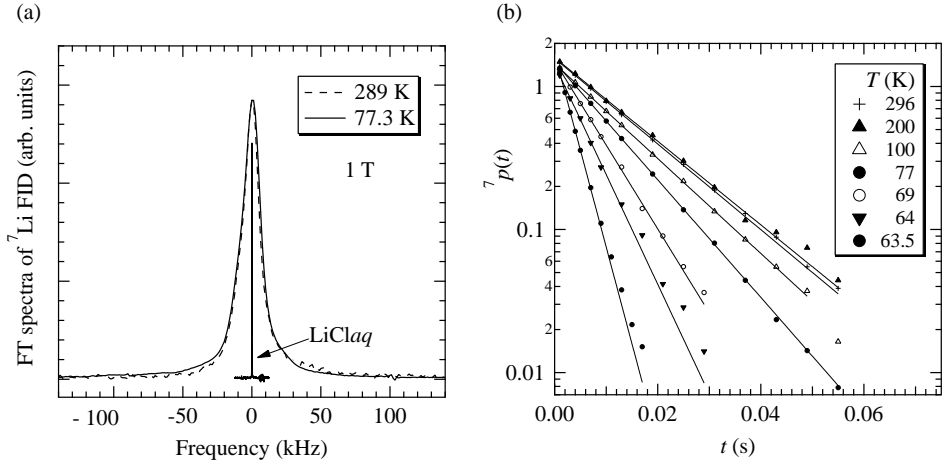}
 \end{center}
 \caption{\label{fig:LiNMR}
(a) Fourier-transformed (FT) spectra of $^7$Li FID signals in LiCrO$_2$ at 77 and 289 K. (b) Recovery curves $^{7}p(t)$ of $^7$Li NMR FID signals. Solid lines are the results from the least-squares fit by a single exponential function
with a time constant $^{7}T_1$.    
 }
 \end{figure}  
 
Figure 2(a) shows the temperature dependence of 1/$^7T_1$. With cooling down, 1/$^7T_1$ shows a critical divergence near $T_\mathrm{N}$, while it levels off at higher temperatures above about 200 K. The high temperature value of 1/$^7T_{1\infty}$ is estimated to be 69 s$^{-1}$. The paramagnetic state above about 200 K is in the exchange narrowing limit. Then, the upper limit of an exchange coupling constant $J$ is $\sim$ 200 K. 
The Curie-Weiss spin susceptibility fit at higher temperatures indicates $J$ = 80 K~\cite{XT,NMR}.  

Figure 2(b) shows normalized (1/$^7T_1$)/(1/$^7T_{1\infty}$) versus reduced temperature $\left| T - T_\mathrm{N}\right|$/$T_\mathrm{N}$. The solid line is the result from the least-squares fit by 1/$T_1$ = ($C/^7T_{1\infty}$)($T/T_\mathrm{N}$ - 1)$^{-w}$ where $C$ and $w$ are the fit parameters. The critical exponent is estimated to be $w$ = 0.45. A mean field theory for a three dimensional isotropic Heisenberg antiferromagnet leads to $w$ = 1/2~\cite{Moriya}. A dynamic scaling theory indicates $w$ = 1/3 for a three dimensional isotropic Heisenberg model~\cite{HH} and $w$ = 2/3 for a three dimensional uniaxial anisotropic Heisenberg model~\cite{RW}. The exponent of $w$ = 0.45 suggests that LiCrO$_2$ in the critical region is described by a three dimensional interaction model. 

\begin{figure}
\begin{center}
\includegraphics[width=14 cm]{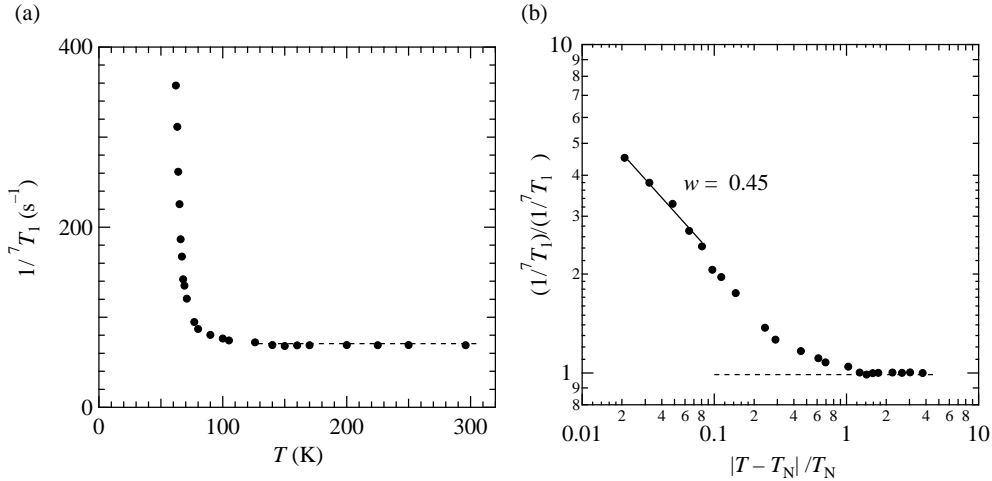}
\end{center}
 \caption{\label{fig:LiT1}
(a) 1/$^7T_1$ against temperature. 1/$^7T_1$ shows a critical divergence near $T_\mathrm{N}$. The dashed line indicates 1/$^7T_{1\infty}$ = 69 s$^{-1}$. (b) Normalized (1/$^7T_1$)/(1/$^7T_{1\infty}$) against reduced temperature $\left| T - T_\mathrm{N}\right|$/$T_\mathrm{N}$. The solid line indicates 1/$T_1$ = ($T/T_\mathrm{N}$ - 1)$^{-w}$. 
}
\label{f2}
\end{figure}

\begin{figure}
\begin{center}
\includegraphics[width=10 cm]{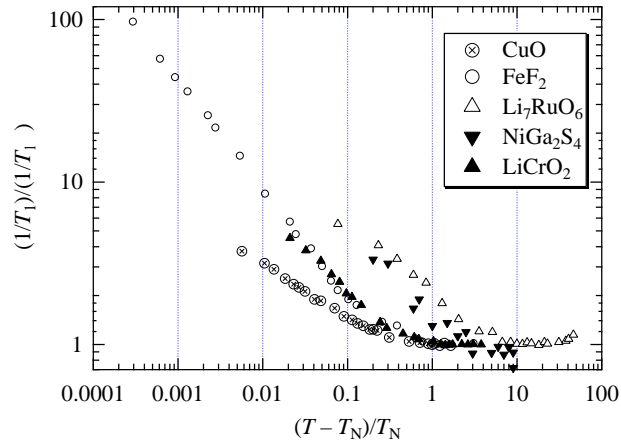}
\end{center}
 \caption{\label{fig:LogLog}
Normalized (1/$T_1$)/(1/$T_{1\infty}$) against reduced temperature ($T - T_\mathrm{N}$)/$T_\mathrm{N}$ for three dimensional (CuO~\cite{CuO}, FeF$_2$~\cite{FeF2}), triangular-lattice (Li$_7$RuO$_6$~\cite{Itoh}, NiGa$_2$S$_4$~\cite{NiGa2S4}) and the present LiCrO$_2$.  
}
\label{f3}
\end{figure}

Figure 3 shows log-log plots of the normalized (1/$T_1$)/(1/$T_{1\infty}$) against the reduced temperature ($T - T_\mathrm{N}$)/$T_\mathrm{N}$ for three dimensional antiferromagnets (CuO~\cite{CuO}, FeF$_2$~\cite{FeF2}), triangular-lattice antiferromagnets (Li$_7$RuO$_6$~\cite{Itoh}, NiGa$_2$S$_4$~\cite{NiGa2S4}) and the present LiCrO$_2$. 
The critical divergence of (1/$T_1$)/(1/$T_{1\infty}$) of LiCrO$_2$ coincides with a part of FeF$_2$, which is a uniaxial anisotropic Heisenberg system.  
The onset of the increase in the NMR relaxation rate near $T_\mathrm{N}$ empirically categorizes the critical region. 
The region of $\left| T - T_\mathrm{N}\right|$/$T_\mathrm{N}\leq$ 10 has been assigned to the renormalized classical regime with the divergent spin-spin correlation length toward $T$ = 0 K~\cite{Itoh}.   
The region of $\left| T - T_\mathrm{N}\right|$/$T_\mathrm{N}\leq$ 1.0 has been assigned to the three dimensional critical regime with the divergent spin-spin correlation length toward $T_\mathrm{N}$. 
Thus, the narrow critical region of $\left| T - T_\mathrm{N}\right|$/$T_\mathrm{N}\leq$ 1 also indicates that LiCrO$_2$ is in the three dimensional critical regime.

\section{Discussions}

The theoretical analysis of the non-linear sigma model for the spin $S$ frustrated quantum antiferromagnets gives us the magnetic correlation length~\cite{The1,The2,The3,The4}
\begin{equation}
\xi\propto\frac{1}{\sqrt{T}} \mathrm{exp}(4\pi\rho_{s}/T)
\label{eq:xiSU2}
\end{equation}
with a spin stiffness constant $\rho_{s}$ and the nuclear spin-lattice relaxation rate
\begin{eqnarray}
\frac{1}{T_1T^3}\propto \mathrm{exp}(4\pi\rho_{s}/T). 
\label{eq:T1SU2} 
\end{eqnarray}  
Here, the spin stiffness constant $\rho_{s}$ is expressed by
\begin{equation}
\rho_{s} = \frac{\sqrt 3}{2}Z_{s}S^2J_s,
\label{eq:RGZ}
\end{equation}
where a renormalization factor $Z_{s}$ is calculated by a spin-wave approximation and 1/$S$ expansion and $J_s$ is the nearest neighbor exchange coupling constant~\cite{The5}. 

\begin{figure}
\begin{center}
\includegraphics[width=14 cm]{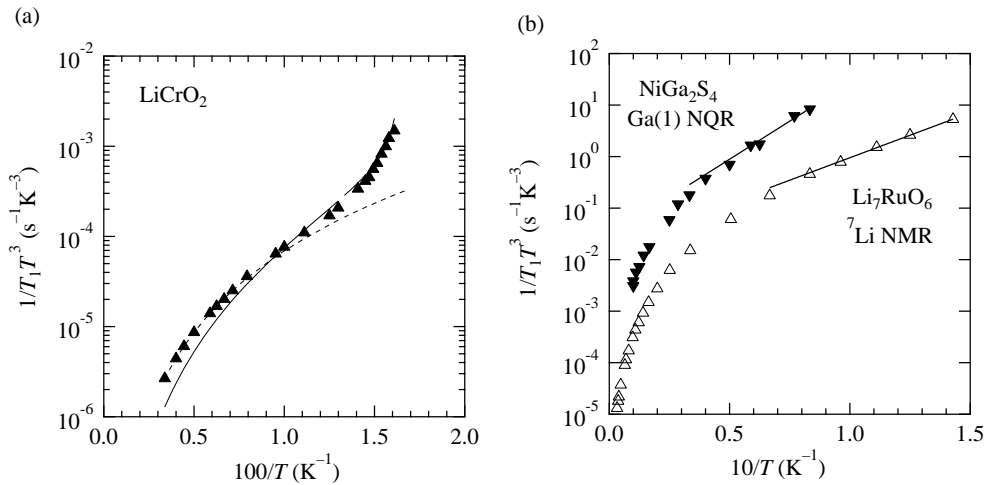}
\end{center}
 \caption{\label{fig:RC}
(a) Semi-logarithmic plot of 1/$^7T_1T^3$ against inverse temperature 100/$T$ for LiCrO$_2$. The dash curve indicates 1/$^7T_1$ = 1/$T_{1\infty}$. The solid curve indicates 1/$^7T_1 \propto (T/T_\mathrm{N} - 1)^{-w}$. 
(b) Semi-logarithmic plots of 1/$T_1T^3$ against inverse temperature 10/$T$ for Li$_7$RuO$_6$~\cite{Itoh} and NiGa$_2$S$_4$~\cite{NiGa2S4}. The solid lines indicate $1/{T_{1}T^{3}}\propto \mathrm{exp}(4\pi\rho_{s}/T)$.  
}
\label{f4}
\end{figure}

\begin{table}
\caption{Spin stiffness constants and exchange coupling constants estimated from the analysis using the two dimensional renormalized classical model for triangular lattice compounds. The data of Li$_7$RuO$_6$ are reproduced from ref.~\cite{Itoh}. The data of NiGa$_2$S$_4$ are estimated from Fig. 4(b) using the experimental $T_1$ values in ref.~\cite{NiGa2S4}. The data of KCrO$_2$ are reproduced from ref.~\cite{KCrO2}.
$J_{\Theta}$ of LiCrO$_2$ is from refs.~\cite{XT,NMR}.
}
\label{t1}
\begin{center}
\begin{tabular}{llll}
\hline
\multicolumn{1}{l}{ } & \multicolumn{1}{c}{4$\pi\rho_s$ (K)} & \multicolumn{1}{c}{$J_s$ (K)}  & \multicolumn{1}{c}{$J_{\Theta}$ (K)}  \\
\hline
Li$_7$RuO$_6$ & 40 & 2.1 & 9.7 \\
NiGa$_2$S$_4$ & 68 & 9.5 & 20 \\
KCrO$_2$ & 130 & 9.3 & 29 \\
LiCrO$_2$ & --- & --- & 80 \\
\hline
\end{tabular}
\end{center}
\end{table}

Figure 4(a) shows semi-logarithmic plot of 1/$^7T_1T^3$ against inverse temperature 100/$T$ for LiCrO$_2$. The dash curve is 1/$^7T_1$ = 1/$T_{1\infty}$. The solid curve is a best fit result of 1/$^7T_1\propto(T/T_\mathrm{N} - 1)^{-w}$ near $T_\mathrm{N}$. No trace of two dimensional renormalized classical regime is found. 

For comparison, Fig. 4(b) shows semi-logarithmic plots of 1/$T_1T^3$ against inverse temperature 10/$T$ for Li$_7$RuO$_6$~\cite{Itoh} and NiGa$_2$S$_4$~\cite{NiGa2S4}. The solid lines are the best fit results of $1/{T_{1}T^{3}}\propto \mathrm{exp}(4\pi\rho_{s}/T)$, which is characteristic of the two dimensional renormalized classical behavior of the triangular lattice spin systems.    

Since the solid lines in Fig. 4(b) well fit the experimental 1/$T_1$ at low temperatures, the spin-spin correlations of Li$_7$RuO$_6$~\cite{Itoh} and NiGa$_2$S$_4$~\cite{NiGa2S4} are in the two dimensional renormalized classical regimes. However, the exchange coupling constants $J_s$'s estimated from eqs. (2) and (3) are smaller than the exchange constants $J_\Theta$'s estimated from the Curie-Weiss susceptibility fit at higher temperatures~\cite{X}. The estimated parameters of 4$\pi\rho_s$, $J_s$ and $J_\Theta$ are listed in Table 1.  
One may find $J_s\sim J_\Theta$/5 for Li$_7$RuO$_6$,  $J_s\sim J_\Theta$/2 for NiGa$_2$S$_4$, and $J_s\sim J_\Theta$/3 for KCrO$_2$.
Since no frustration effects are taken into consideration in eq. (3), the reason of $J_s < J_\Theta$ can be traced back to the spin frustration effects on a spin-spin correlation function at a low frequency. Actually, the reduction of the spin stiffness constant $\rho_s$ due to the spin frustration (Z$_2$ vortices) is seen in the numerical studies of Heisenberg frustrated spin systems~\cite{SF}.  

\section{Conclusions}

We have made a detailed experimental study of the $^7$Li NMR FID signal in the paramagnetic state of the triangular lattice antiferromagnet LiCrO$_2$. The critical behavior of the $^7$Li nuclear spin-lattice relaxation rate 1/$T_1$ of the FID signal near $T_\mathrm{N}$ is found to be well described by a power law. The critical exponent takes $w$ = 0.45. The narrow critical region and no trace of two dimensional renormalized classical regime are found. 
The three dimensional exchange interaction may play a central role in the critical behavior of LiCrO$_2$.

\end{document}